\documentclass[aps,prb,twocolumn, 10pt,tightenlines,amsmath,amssymb]{revtex4}
\usepackage{graphicx}
\usepackage{dcolumn}
\usepackage{bm}
\newcommand{\beq}{\begin{equation}}
\newcommand{\eeq}{\end{equation}}

\def\nothing#1{}
\begin{document}

\title{Injectance and a paradox}
\author{Urbashi Satpathi}
\affiliation{Unit for Nano Science and Technology, 
S. N. Bose National Centre for Basic Sciences,
Block JD, Sector III, Salt Lake, Kolkata 700 098, India.}
\email{urbashi@boson.bose.res.in}   
\author{P Singha Deo}
\affiliation{Unit for Nano Science and Technology,
S. N. Bose National Centre for Basic Sciences, 
Block JD, Sector III, Salt Lake, Kolkata 700 098, India.}
\email{deo@bose.res.in}
\date{\today}

\begin{abstract}
Quantum mechanics manifests in experimental observations in several ways. Hauge \textit{et al.} (1987) and Leavens \textit{et al.} (1989) had pointed out that interference effects dominate a physical quantity called injectance. We show that, very paradoxically, the interference related term vanish in a quantum regime making semi-classical formula for injectance exact in this regime. This can have useful implications to experimentalists as semi-classical formulas are much more simple. There are other puzzling facts in this regime like an ensemble of particles can be transmitted without any time delay or negative time delays, whereas the reflected particles are associated with a time delay.
\end{abstract}

\maketitle

A series of experiments has recently confirmed that scattering phase shifts in quantum mechanics can be measured \cite{hei00,sch,kob1,kob2} . Non-locality in quantum mechanics does not allow us to determine a particular path in which the electron wave propagates. This is unlike classical waves. This problem was overcome by using additional probes and controlled decoherence \cite{sch}. Ref. [\onlinecite{hilbertTransform}] confirms that Hilbert  transform of the measured conductance data gives the measured phase data which confirms that the scattering phase shift was correctly measured.
Thus the scattering phase shift as well as the scattering cross section of an arbitrary quantum system (say, its impurity configuration and confinement potential is not known) can be measured. While measuring scattering cross section is an old story, measuring scattering phase shifts is novel and new. So one can ask the question that from the measured scattering phase shift what can we learn about the quantum system. 

Friedel sum rule (FSR) relates scattering phase shift to density of states (DOS) in the system and Wigner Smith delay time  (WSDT) relates scattering phase shift to a time scale at the resonances. The two are basically the same as both density of states and life time of a resonance are given by the imaginary part of the retarded Greens function. Although, it is to be noted that different works have interpreted the WSDT in different ways which will be discussed later. They (FSR and WSDT) are semi-classical formulas that are not a priori applicable to quantum systems in mesoscopic regime. The quantum versions of these formulas are not completely in terms of the experimental data. A number of works has studied FSR and WSDT in the single channel case \cite{jpcm,lee,yey,tan}. Ref. [\onlinecite{jpcm}] has shown that in the single propagating channel regime, for any arbitrary potential that has symmetry in $x$-direction i.e., $V(x,y)=V(-x,y)$, FSR and WSDT are exact at the resonances of the system, essentially due to the fact that the resonances are Fano resonances.
Another physical quantity of interest is injectivity or injectance \cite{btp}. In this work we calculate injectivity and injectance in the single channel as well as two channel regime, as this also reveals the paradoxical nature of the scattering phase shift at Fano resonance.

Let us consider plane wave incident from left hand side on a three dimensional scatterer. The scattered or asymptotic wave function in spherical polar coordinates (r, $\theta$, $\phi$) is given by \cite{merz},
$$\frac{1}{r}sin(kr-\frac{l\pi}{2}+\theta_t) P_l cos(\phi)$$
Here $\theta_t$ is the scattering phase shift. 
Assume a large sphere of radius R and let us count the number of nodes inside the sphere. Number of states is one more than the number of nodes (number of nodes +1). To count the number of nodes inside the sphere we can set the wave function to zero on the boundary of the sphere, i.e.,\cite{ziman}
\begin{equation}
kR-\frac{l\pi}{2}+\theta_t=n\pi \  \ \ \ or,\frac{d\theta_t}{dk} + R = \frac{dn}{dk} \pi       \label{five}
\end{equation}
In absence of scatterer,
\begin{equation}
 kR =n_0\pi \ \ \ \ or,R=\frac{dn_0}{dk}\pi                 \label{six}
\end{equation}
As, $E=\frac{\hbar^2k^2}{2m_e}$, we get from (\ref{five}) and (\ref{six}),
$\frac{d\theta_t}{dE}=\pi[\rho(E) - \rho_0(E)]$
where, $\rho(E)=\frac{dn}{dE}$ is the density of states in the presence of scatterer and $\rho_0(E)=\frac{dn_0}{dE}$ is the density of states in the absence of scatterer. Instead one can put $\theta_t+kR\equiv\theta_t$ that is scattering phase shift is defined with respect to phase shift $kR$ in absence of scatterer. In which case FSR becomes \cite{ziman},
\begin{equation}
\frac{d\theta_t}{dE}=\pi\rho(E)     \label{three}
\end{equation}
In the rest of the paper we will use this definition for scattering phase shift.
This is a semi-classical result because it is valid only when, $dn\ll n$. This assumption is related to non-dispersive wave packets or stationary phase approximation as shown below. In this case we restrict to one dimension as Eq. (\ref{three}) can be shown in one dimension too where the large sphere of radius R become just two points at a large distance of $x=\pm R$. 

Consider a wave packet incident from left on a one dimensional scatterer, i.e.,
$\psi_{in}(x,t)=\int_{-\infty}^{+\infty}a_k e^{ikx-i\omega t} dk$.
After scattering the transmitted wave packet at $(x+\Delta x , t+\Delta t)$ is,
$$\psi_{sc}(x+\Delta x,t+\Delta t)=$$
\begin{equation}
\int_{-\infty}^{+\infty} \mid t(k)\mid a_k e^{ik(x+\Delta x)-i\omega(t+\Delta t)+i\theta_t} dk
\end{equation}
where, $t(k)$ is the transmission amplitude and $\theta_t$ is its phase. A classical particle is either transmitted completely or reflected completely without any distortion.
Assuming there is no reflected part, along with the assumption of no dispersion of wave packet (known as stationary phase approximation), we can write,
\begin{equation}
kx+k\Delta x-\omega t-\omega\Delta t+\theta_t= K   \label{one}
\end{equation}
where K is a constant. This implies that the phase of the amplitude component $a_k$ remain stationary or that the wave packet remains undispersed. Therefore from (\ref{one}),
\begin{figure}
\centering
{\includegraphics[width=6cm,keepaspectratio]{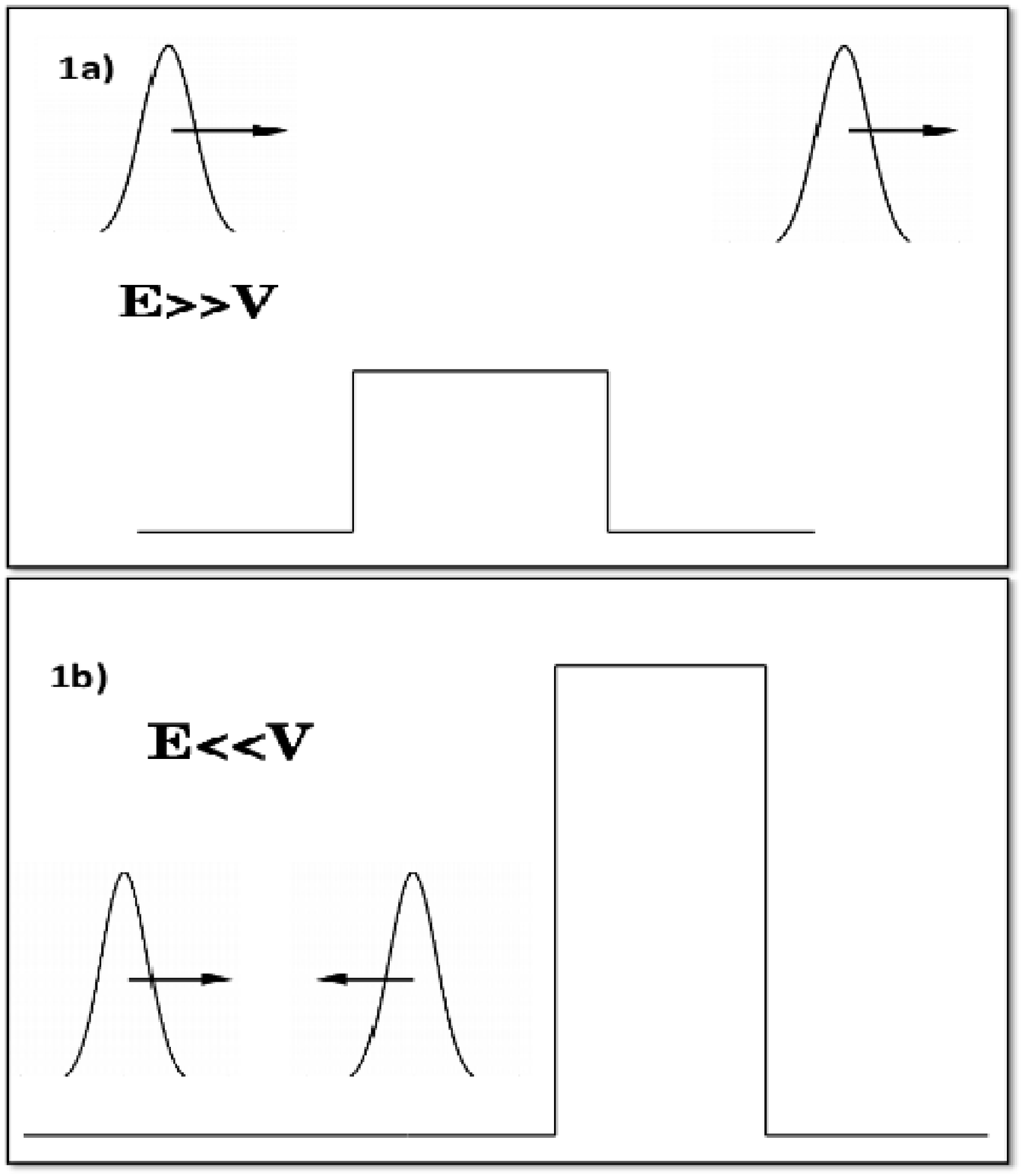}}
\caption{\label{fig1}
The figure shows a wave packet is incident on a barrier of height $V$ in one dimension.
a)The energy of incidence E of the centroid of the wave packet is much larger than the barrier potential V, i.e $E\gg V$, and, b) The energy of incidence E is much smaller than the barrier potential V, i.e $E\ll V$. These limits correspond to semi-classical regimes.}
\end{figure}
\begin{equation}
\frac{d\theta_t}{d\omega} = \Delta t -\frac{\Delta x}{v_g}\   \  \  \  \  \ or, \ \ \   \hbar\frac{d\theta_t}{dE}= \Delta t -\Delta t_0               \label{seven}
\end{equation}
where $v_g=\frac{d\omega}{dk}$ is the group velocity.
Semi-classical behavior is pictorially depicted in figures 1a) and 1b). When $ E\gg V$ then a wave packet  can get completely transmitted with Eq. (\ref{one}) approximately satisfied (it is never exactly satisfied as electron dispersion is quadratic). It can also happen when $E\ll V$ then a wave packet can get completely reflected with very little dispersion. In this case, $\theta_t$ is to be replaced by $\theta_r$, the reflected phase shift. Thus we find that when a wave packet remains undispersed, then the time spent by the particle in the scattered region ($\Delta t -\Delta t_0$), is given by energy derivative of scattering phase shift. This life time (WSDT) is also related to DOS as both are given by the imaginary part of retarded Green's function [ref. [\onlinecite{dat}] page 155-156]. Hence, as we have seen in Eq. (\ref{three}), the DOS can also be found from the energy derivative of scattering phase shift, which is FSR. These formulas are obviously not valid in the quantum regime i.e., $E\sim V$, where there will be transmission as well as reflection. Also in this regime the scattering phase shifts will be so strongly energy dependent that stationary phase approximation cannot remain valid and a wave packet will always undergo dispersion. Refs. [\onlinecite{hff,lev}] have looked into the one dimensional problem when both transmission and reflection is present. They have computed the correction term and explained the significance of it. They have shown that the correction term arises because of quantum interference effect. Note that both dispersion and quantum interference arises due to the superposition principle in quantum mechanics. Their result will be discussed latter.

 The quantum regime can be treated exactly and also yields the following formula for time known as Larmor precession time (LPT) \cite{btp},
\begin{equation}
\tau(\alpha , r , \beta )= -\frac {\hbar}{4\pi i} \left[ s_{\alpha\beta} ^\dagger \frac {\Delta s_{\alpha\beta}}{e \Delta \textit{U}(r)} - \frac {\Delta s_{\alpha\beta} ^\dagger}{e\Delta \textit{U}(r)}s_{\alpha\beta} \right] \label{two}
\end{equation}
Connection between LPT and WSDT or FSR is explained below. $\tau(\alpha , r , \beta )$ is the time spent by one particle incident along channel $\beta$ and scattered to channel $\alpha$ at r. LPT is exact in the sense that when summed over $\alpha$ and $\beta$ and divided by $\hbar$, it gives the exact DOS as calculated from the internal wave function. $\Delta$ stands for functional derivative. Note that the functional derivative is with respect to the local potential implying that exact DOS cannot be expressed using asymptotic wave function alone. Dividing  Eq. (\ref{two}) by $\hbar$ we get a partial local density of states (PLDOS) for such a process (a particle incident along $\beta$  and scattered to $\alpha$).
That is $\frac{\tau(\alpha , r , \beta)}{\hbar}=\nu(\alpha,r,\beta)$, where $\nu(\alpha,r,\beta)$ is partial local density of states. Such a process requires to specify both the incoming and outgoing channel which is impossible in quantum mechanics because a single particle in quantum mechanics behave probabilistically when it encounters a potential and gets scattered. Only when an ensemble of particles is considered, the probability of transmission and that of reflection are given by Schr\"{o}dinger equation. One can indirectly measure the consequences of such a PLDOS \cite{btp}, but an experimental set up that can directly probe this PLDOS requires us to take a sum over at least one of the channels (i.e., $\alpha$ or $ \beta $). So, we can specify the incoming channel and scattering can take the particle to any arbitrary output channel. That is, $ \sum_ {\alpha} \nu(\alpha,r,\beta) = I(r,\beta)$ is a physical quantity called injectivity.
\begin{eqnarray}
I(r,\beta) & = & \sum_{\alpha} -\frac {1}{4\pi i} \left[ s_{\alpha\beta} ^\dagger \frac {\Delta s_{\alpha\beta}}{e \Delta \textit{U}(r)} - \frac {\Delta s_{\alpha\beta} ^\dagger}{e\Delta \textit{U}(r)}s_{\alpha\beta} \right]  \nonumber\\
& = & \sum_{\alpha} -\frac {1}{2\pi} \mid s_{\alpha\beta}\mid^2\frac{\Delta \theta_{s_{\alpha\beta}}}{e\Delta \textit{U}(r)}   			\label{i(r,b)}
\end{eqnarray}
where, $\theta_{s_{\alpha\beta}} = \arctan \frac{Im(s_{\alpha\beta})}{Re(s_{\alpha\beta})}$.
This quantity can be experimentally observed and is the topic of study in this work. A typical situation where this quantity can be observed is when we bring a scanning tunneling microscope (STM) tip close to a mesoscopic sample at $r$ connected to one or more leads. Injectivity gives current delivered by the tip. 

When injectivity is summed over $\beta$ then we get DOS which has been studied earlier \cite{jpcm}. Not summing over $\beta$ reveals the true nature of the paradox. This is because then the incident wave packet comes along a single channel $\beta$ and after scattering it will either disperse or will not disperse. If we sum over $\beta$ then the incidence is along all possible channels. So wave packets are incident along $\beta$ channels and their scattering and dispersion can compensate each other. Also refs. [\onlinecite{hff,lev}] have calculated the correction term for this injectivity in the single channel case and the paradox can be quantitatively explained in terms of this correction term. 
The semi-classical limit can be obtained from the following substitution,
\begin{equation}
-  \int _\Omega d^3 r\frac{\Delta}{e\Delta \textit{U}(r)} \longrightarrow \frac{d}{d\textit{E}}     \label{intOmega}
\end{equation}
Hence,
\begin{equation}
I(E) \approx \sum_{\alpha} \frac {1}{2\pi} \mid s_{\alpha\beta}\mid^2\frac{d}{dE}\theta_{s_{\alpha\beta}}            \label{i(b,E)}
\end{equation}
This quantity is called injectance. $\mid s_{\alpha\beta}\mid^2$ appears because when we allow scattering (Eq. (\ref{seven}) correspond to a case when everything is transmitted without scattering) then $\mid s_{\alpha\beta}\mid^2$ number of particles are scattered from channel $\beta$ to channel $\alpha$ and each ones contribution to time is related to $\frac{d\theta_{s_{\alpha\beta}}}{dE}$. 
In the single channel case, it can be explicitly written as
\begin{equation}
I(E)\approx  \frac {1}{2\pi}\left[ \mid r\mid^2 \frac{d\theta_{r}}{dE}+\mid t\mid^2 \frac{d\theta_{t}}{dE}	\right] 				\label{I(E)}
\end{equation}
The approximate equality can be replaced by an equality if we add a correction term on the right hand side \cite{hff,lev}.
\begin{equation}
I(E)= \frac {1}{2\pi} \left[ \mid r\mid^2 \frac{d\theta_{r}}{dE}+\mid t\mid^2 \frac{d\theta_{t}}{dE}+\frac{m_e\mid r\mid}{\hbar k^2} \sin (\theta_{r})\right] 	  \label{I(crr)}
\end{equation}
where $k$ is the wave vector of the incident channel. Refs. [\onlinecite{hff,lev}] have stressed that the correction term can be zero if $\mid r\mid=0$ (corresponding to Fig.1a)), but $\sin (\theta_{r})$ cannot be zero as it comes from quantum interference which is always there in a quantum system. $\sin (\theta_{r})$ can be zero only when the potential is infinitely high because then there is obviously no interference effect \cite{hff,lev} (corresponding to Fig.1b)). Or it sometimes become zero far away from resonances that are almost classical regimes \cite{hff,lev}. Therefore, just as dispersion of a wave packet cannot disappear, interference effects cannot disappear. Both originate from the linear superposition principle in quantum mechanics. Only when dispersion or interference can be ignored we get the semi-classical limit, i.e, Eq. (\ref{i(b,E)}).
\begin{figure}
\centering
{\includegraphics[width=6cm,keepaspectratio]{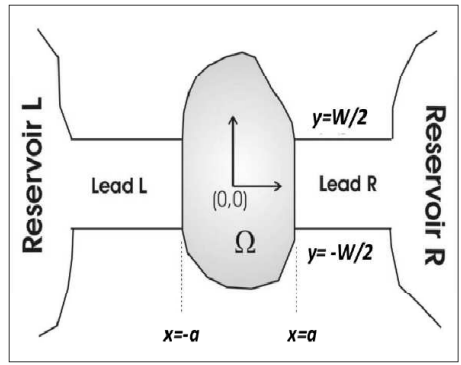}}
\caption[]{\label{fig2}
A general scattering problem in quasi one dimension. The reservoir L inject electrons to the left lead L and hence to the scattering region with arbitrary potential $V_g(x,y)$, shown by the shaded region. Reservoir R absorb electrons transmitted through the shaded region, through right lead R. The electrons reflected from the shaded region are collected in reservoir L. Here we have considered one propagating channel in the leads and electrons are incident from the left.}
\end{figure}

We will first show that there is a situation where $\sin(\theta_r)$ can become zero at a resonance where $\mid r\mid\neq 0$, making (\ref{i(b,E)}) exact in a quantum regime. This is in complete contradiction to what is known so far \cite{hff,lev}. A general proof for a single channel scattering is given below. This can have useful implications to experimentalists in the sense that although injectance depends on the potential inside the scatterer, it can be determined from asymptotic wave function that can be experimentally measured. Consider a quasi one dimensional (Q1D) system with scattering potential $V_g(x,y)$ shown in Fig.2 by the shaded region. Lead L and lead R connect the system to electron reservoirs L and R, respectively. They act as source and sink for electrons, respectively.
The confinement potential in the leads, in the $y$-direction (or, transverse direction) is taken to be hard wall, given by,
$$V(y) = \infty   \,\, \textrm{for $ \mid y \mid \geq\frac{W}{2}$}  $$
$$ = 0     \,\, \textrm{for $\mid y \mid < \frac{W}{2}$}  $$
The Schr\"{o}dinger equation in the two dimensional system (the third direction is eliminated by quantization \cite{dat}) is,
$$\left[ -\frac{\hbar^2}{2m_e}\left( \frac{\partial^2}{\partial x^2}+\frac{\partial^2}{\partial y^2}\right)+V(y)+V_{g}(x,y)\right]\Psi(x,y)$$
\begin{equation}
=E\Psi(x,y) 
\end{equation}
where $m_e$ is the mass of the electron, $W$ is the width of leads L and R.
In the leads where there is no scatterer ($V_g(x,y)=0$), the Schr\"{o}dinger equation can be decoupled. The $y$-component is,
\begin{equation}
\left[ -\frac{\hbar^2}{2m_e}\frac{d^2}{dy^2}+V(y)\right]\chi_m(y)=\varepsilon_m\chi_m(y)
\end{equation}
and the $x$-component is,
\begin{equation}
-\frac{\hbar^2}{2m_e}\frac{d^2}{dx^2}c_m(x)=(E-\varepsilon_m)c_m(x)					\label{E}
\end{equation}
with $\chi_m(y)=\sqrt\frac{2}{W}\sin\frac{m\pi}{W}(y+\frac{W}{2})$ and $\varepsilon_m=\frac{m^2 \pi^2 \hbar^2}{2 m_e W^2}$. 
$E$ is the energy of incidence from reservoir L, given by,
\begin{equation}
E=E_{m,k_m}=\displaystyle \frac{m^2\pi^2\hbar ^2}{2m_eW^2}+ \frac{\hbar ^2k_m^2}{2m_e}						\label{E_m}
\end{equation}
It is known that for potentials that have symmetry in $x$-direction, i.e.,$V(x,y)=V(-x,y)$, we can write solutions to (\ref{E}) given below.
\begin{equation}
c_m^{e}(x)=\sum_{m=1}^{\infty}(\delta_{mn} e^{-ik_{m}x} - S_{mn}^{e} e^{ik_{m}x}) \frac{1}{\sqrt k_{m}}               \label{psi_e}
\end{equation}
\begin{equation}
c_m^{o}(x)=\sum_{m=1}^{\infty}(\delta_{mn} e^{-ik_{m}x} - S_{mn}^{o} e^{ik_{m}x}) \frac{1}{\sqrt k_{m}}               \label{psi_o}
\end{equation}
where, $c_m^{e}(x)=c_m^{e}(-x)$ and $c_m^{o}(x)=- c_m^{o}(-x)$. Then both transmitted wave function (at $x>a$) and reflected wave function (at $x<-a$) are given by \cite{jpcm,bay}
\begin{equation}
c_m(x)=\frac{c_m^{e}(x)-c_m^{o}(x)}{2}		\label{cm}
\end{equation}

$E$ can be so adjusted by adjusting the Fermi energy of reservoir L, that $\frac{\pi^2 \hbar^2}{2 m_e W^2} < E < \frac{4 \pi^2 \hbar^2}{2 m_e W^2}$.  
Then $k_1$ is real and from (\ref{psi_e}), (\ref{psi_o}) and (\ref{cm})
$$c_1(x)=e^{ik_1x} + {\tilde r_{11}} e^{-ik_1x} \,\, \textrm{for $x<-a$} $$
$$= {\tilde t_{11}} e^{ik_1x} \,\,\textrm{for $x>a$}$$
where
\begin{equation}
{\tilde r_{11}}= -\frac{(S_{11}^{o} +S_{11}^{e})}{2}										\label{rmn}
\end{equation}
\begin{equation}
{\tilde t_{11}}= \frac{(S_{11}^{o} -S_{11}^{e})}{2}											\label{tmn}
\end{equation}
For $m>1$, energy conservation in (\ref{E_m}) is not violated as $k_m^2$ can become negative. That yields evanescent solutions with $k_m\rightarrow i\kappa_m$. Inclusion of the evanescent modes is very important to get the correct solutions.
Due to the same principle, i.e., any function can be written as a sum of an even function and odd function and any square matrix can be written as a sum of a symmetric matrix and an anti-symmetric matrix, the wave function in the scattering region ($-a<x<a$) can be written as a sum of an even function and an odd function. We denote them as $\eta_{m}^{e}(x,y)$ and $\eta_{m}^{o}(x,y)$, i.e., $\Psi(x,y)=\frac{\eta_{m}^{e}(x,y)+\eta_{m}^{o}(x,y)}{2}$ for $-a< x< a$.
\begin{equation}
\eta_{m}^{e}(x,y)=\sum_{n=1}^{\infty}d_{n}\zeta_{n}^{e}(x,y)					\label{eta_e}
\end{equation}
\begin{equation}
\eta_{m}^{o}(x,y)=\sum_{n=1}^{\infty}d_{n}\zeta_{n}^{o}(x,y)					\label{eta_o}
\end{equation}
$\zeta_{n}^{e}$ and $\zeta_{n}^{o}$ are the basis states that satisfy the condition that they go to zero at the upper edge and lower edge of the shaded region in Fig.2.

One can define the matrix elements,
\begin{equation}
F_{m,n}^{eo}=\frac{2}{W(k_m k_n)^\frac{1}{2}}\int_{-b}^{b}\chi_{m}(y)\left(\frac{\partial \zeta_{n}^{eo}}{\partial x} \right)_{x=a} dy		\label{Fmn} 
\end{equation}
Here $eo$ stands for even or odd, i.e., $e/o$.
One can match the wave function and conserve the current at $x=\pm a$ for all $y$ to obtain a matrix equation given by,
\begin{equation}
\sum_{n=1}^{\infty}\left[ F_{rn}^{eo}-i\delta_{rn}\right] e^{ik_{n}a} S_{nm}^{eo}=	\left[ F_{rm}^{eo}+i\delta_{rm}\right] e^{-ik_{m}a} 			\label{Frn}
\end{equation}
Solving for $S_{mn}^{eo}$, we can find the scattering matrix elements.
Bound states can be determined from the singularities of the matrix equation (\ref{Frn}), on setting right hand side to zero. That is
\begin{equation}
det\left[ F_{cc}^{eo} -i 1\right] =0																				\label{Fcc}
\end{equation}
Here `$cc$' means evanescent channel (or closed channel) for which both $k_m$ and $k_n$ in (\ref{Fmn}, \ref{Frn}), are imaginary.
Solving Eq. (\ref{Frn}), one can get \cite{jpcm,bay},
\begin{equation}
S_{11}^{eo} = e^{-2ik_{1}a}\frac{G^{eo} + i}{G^{eo} - i}= e^{2i arccot[G^{eo}]}= e^{2i\theta^{eo}}					\label{S11}
\end{equation}
where,
\begin{equation}
G^{eo} =F_{11}^{eo} -\sum_{m=2,n=2} F_{1n}^{eo} \left[ \left( F_{cc}^{eo} - i 1\right)^{-1}\right] _{nm} F_{m1}^{eo} 				\label{G11}
\end{equation}
and
\begin{equation}
\theta^{eo}= arccot[G^{eo}]														\label{theta_eo}
\end{equation}
Here scattering phase shift $\theta^{eo}$ is defined with respect to phase shift in the absence of scatterer.
Putting (\ref{S11}) in (\ref{rmn}), (\ref{tmn}) and defining new variables,
\begin{equation}
\phi=\theta^{e}-\theta^{o} 		\,\, \textrm{and \;  $\theta_r=\theta^{e}+\theta^{o}$}		\label{new1}
\end{equation} 
we get  transmission amplitude and reflection amplitude
\begin{equation}
{\tilde t_{11}}=i\sin(\phi)e^{i\theta_r} 		 \,\, \textrm{and \;  ${\tilde r_{11}}=\cos(\phi)e^{i\theta_r}$}		\label{new2}
\end{equation}
The correction term in (\ref{I(crr)}) is $\frac{m_e\mid {\tilde r_{11}}\mid}{\hbar k_1^2} \sin(\theta_r)$.

Threshold energy $E$ for the second channel is given by $E= \frac{4 \pi^2 \hbar^2}{2 m_e W^2}$, i.e., above this energy $k_2$ becomes real. Below this energy the second channel can have bound states. Such bound states will occur at energies given by the solution to Eq. (\ref{Fcc}). At these energies the first channel can be propagating as its threshold is given by $E= \frac{\pi^2 \hbar^2}{2 m_e W^2}$ and $S_{11}$ is given by Eq. (\ref{S11}). But at bound state energy, $G^{eo}$ will diverge as it includes matrix elements of $\left[ F_{cc}^{eo} -i 1\right]^{-1}$ as can be seen from equations (\ref{Fcc}) and (\ref{G11}). That in turn implies that at resonance (as can be seen from Eq. (\ref{theta_eo}))
\begin{equation}
\theta^{e}=p \pi        \,\, \textrm{and \;   $\theta^{o}=q \pi$}							\label{theta1}
\end{equation}
Therefore, from Eq. (\ref{theta1}), 
\begin{equation}
\sin(\theta_r)=\sin(\theta^e+\theta^o)=0		\label{new3}
\end{equation}
Thus we have shown that the correction term in Eq. (\ref{I(crr)}) is zero precisely because $\sin(\theta_r)=0$ but $\mid r\mid\neq0$ (\ref{new1}, \ref{new2}). Note that if we ignore the correction term, then all terms on right hand side can be determined experimentally by measuring asymptotic solutions. So injectance can be known from the asymptotic solutions. So the correction term being zero at a resonance can have useful implication to experimentalists.
\begin{figure}
\centering
{\includegraphics[width=6cm,keepaspectratio]{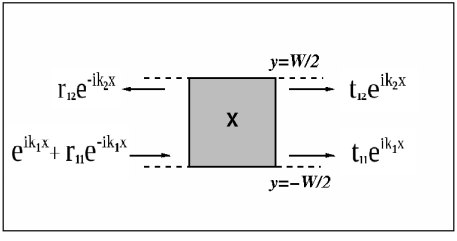}}
\caption[]{\label{fig3}
A similar scattering set up as in Fig.2, with general potential $V_g(x,y)$ replaced by a delta function potential $\gamma \delta(x)\delta(y-y_j)$. Here we are considering two propagating channels. Incidence is along only one channel which is the fundamental channel ($m=1$) from the left.}
\end{figure}

We cannot study the multichannel case generally. We will study the two channel case for a particular potential, i.e., $V_g(x,y)=\gamma \delta(x)\delta(y-y_j)$, to show the paradox there along with other puzzling facts.
In Fig 3, the shaded region is a two dimensional quantum wire with a delta function potential at position (0, $y_j$) marked X. The dotted lines represent the fact that the quantum wire is connected to electron reservoirs.  

Injectance can be obtained from the scattering (S) matrix as described earlier, for the system described in Fig 3. The two channel injectance with incidence along $m=1$ channel from left as shown in Fig.3, in the semi-classical limit is given by,

\begin{eqnarray}
I_1(E) & \approx & \frac{1}{2\pi} \left[ \mid r_{11}\mid^2 \frac{d\theta_{r_{11}}}{dE}+\mid r_{12}\mid^2 \frac{d\theta_{r_{12}}}{dE}\right.  \nonumber\\
& + &\left.\mid t_{11}\mid^2 \frac{d\theta_{t_{11}}}{dE}+\mid t_{12}\mid^2 \frac{d\theta_{t_{12}}}{dE}\right] 						\label{eq1}
\end{eqnarray}
Here, the subscript $1$ signifies the incident channel transverse quantum number, i.e., $m=1$.
We can break this up as
$I_1(E)\approx I_1^L(E)+I_1^R(E)$,
where, 
\begin{equation}
I_1^L(E)=\frac{1}{2\pi}\left\lbrace \mid r_{11}\mid^2\frac{d\theta_{r_{11}}}{dE}+\mid r_{12}\mid^2\frac{d\theta_{r_{12}}}{dE}\right\rbrace
\end{equation}
and 
\begin{equation}
I_1^R(E)=\frac{1}{2\pi}\left\lbrace \mid t _{11}\mid^2\frac{d\theta_{t_{11}}}{dE}+\mid t _{12}\mid^2\frac{d\theta_{t_{12}}}{dE}\right\rbrace	   \label{new8}
\end{equation}
That is, $I_1^L(E)$ consist of reflection channels and $I_1^R(E)$ consist of transmission channels.
The correction term depends on parameters of the incident channel only and gives the following identity.
\begin{equation}
I_1(E)=I_1^L(E)+I_1^R(E)+\frac{1}{2\pi}\frac{m_e \mid r_{11}\mid}{\hbar k_1^2} \sin (\theta_{r_{11}})
\end{equation}
In the similar way, injectance of channel 2 is,
\begin{eqnarray}
I_2(E) &= & \frac{1}{2\pi} \left[ \mid r_{22}\mid^2 \frac{d\theta_{r_{22}}}{dE}+\mid r_{21}\mid^2 \frac{d\theta_{r_{21}}}{dE}+\mid t_{22}\mid^2 \frac{d\theta_{t_{22}}}{dE}\right. \nonumber\\
& + &\left.\mid t_{21}\mid^2 \frac{d\theta_{t_{21}}}{dE} +\frac{m_e \mid r_{22}\mid}{\hbar k_2^2} \sin (\theta_{r_{22}})\right] \label{new6}
\end{eqnarray}
and in the semi-classical limit, it is given by,
\begin{eqnarray}
I_2(E) &\approx &\frac{1}{2\pi} \left[ \mid r_{22}\mid^2 \frac{d\theta_{r_{22}}}{dE}+\mid r_{21}\mid^2 \frac{d\theta_{r_{21}}}{dE}\right.  \nonumber\\
& + &\left.\mid t_{22}\mid^2 \frac{d\theta_{t_{22}}}{dE}+\mid t_{21}\mid^2 \frac{d\theta_{t_{21}}}{dE}\right]	\label{new7}
\end{eqnarray}
We have verified this numerically. The RHS can be determined from the S-matrix. This S-matrix approach is easily accessible to experimentalists. We will now calculate this injectance from internal wave function which allows numerical verification.
\begin{equation}
I_1(E)=\int_{-\infty}^\infty dx \int_{-\frac{W}{2}}^{\frac{W}{2}}dy \sum_{k_1}\left|\Psi(x,y,1)\right|^2 \delta (E-E_{1,k_1}) 				\label{del}
\end{equation}
where $\Psi(x,y,1)$ is the wave function in the scattering region for incidence along channel $m=1$, and is of the form \cite{bag},
\begin{equation}
\Psi(x,y,1)=\sum_m f_m(x,1) \chi_m(y)				\label{psi1}
\end{equation}
Here $\chi_m(y)s$ are solutions in the leads in the transverse direction which is a square well potential in $y$-direction. $\chi_m(y)s$ form a complete set and (\ref{psi1}) is derived from the fact that at a given point $x$, the wave function in the scattering region can be expanded in terms of $\chi_m(y)s$. $f_m s $ are generally of the form given below \cite{bag},
$$f_1(x,1)= e^{ik_1x} + r_{11} e^{-ik_1x} \,\, \textrm{for $x<0$}$$
$$= t_{11} e^{ik_1x} \,\,\textrm{for $x>0$} $$
$$f_2(x,1)=r_{12} e^{-ik_2x} \,\, \textrm{for $x<0$} $$
$$= t_{12} e^{ik_2x} \,\,\textrm{for $x>0$} $$
and for $m>2$,
$$f_m(x,1)=r_{1m} e^{\kappa_m x} \,\, \textrm{for $x<0$} $$
$$= t_{1m} e^{-\kappa_m x} \,\,\textrm{for $x>0$} $$
Here $r_{mn}$ and $t_{mn}$ are unknowns to be determined. The scattering problem described above can be solved using mode matching technique \cite{bag}. 
The reflection amplitudes are given by,
\begin{equation}
r_{mn}(E) =
-\frac{i\frac{\Gamma_{mn}}{2\sqrt{k_{m}k_{n}}}}{1+\sum _{e}\frac{\Gamma_{ee}}{2\kappa_{e}}+
i\sum _{m}\frac{\Gamma_{mm}}{2k_{m}}}	\label{new9}
\end{equation}
$\Gamma_{mn}$ is the coupling strength between $m^{th}$ and $n^{th}$ modes, given by
$\Gamma_{mn}=\gamma \sin\frac{m\pi}{W}(y_j+\frac{W}{2})\sin\frac{n\pi}{W}(y_j+\frac{W}{2})$.
The transmission amplitudes are given by,
$t_{mn}(E) =r_{mn}(E)$ for $m\neq n$  and, $t_{mm}(E) = 1+r_{mm}(E)$.
$\sum_e$ denotes sum over evanescent modes and runs from 3 to $\infty$, while $\sum_m$ denotes the same for propagating modes (i.e. m=1 and m=2). 
$\theta_{r_{mn}}=\arctan \frac{Im  (r_{mn})}{Re ( r_{mn})}$. We will present our results for the case  of  two propagating channels but the analysis and results are same for any number of channels. 

If the delta function potential is negative ($\gamma<0$) then there can be bound states \cite{bag}. In general, the quasi bound state for channel m=s is given by,
\begin{equation}
1+\sum _{e=s}^{\infty}\frac{\Gamma_{ee}}{2\kappa_{e}} = 0                                            \label{bound}
\end{equation}
Only the bound state for m=1 channel is a true bound state and it is given by the solution to the following equation,
$1+\sum _{e=1}^{\infty}\frac{\Gamma_{ee}}{2\kappa_{e}} = 0$. The bound state for m=2 is given by, $1+\sum _{e=2}^{\infty}\frac{\Gamma_{ee}}{2\kappa_{e}} = 0$.
At this energy we get a bound state for m=2, but at that energy m=1 channel is a propagating channel. Hence the bound state given by this equation is a quasi bound state or a resonance.    

The delta function in Eq. (\ref{del}) summed over $k_1$ can be shown to yield a factor $\frac{1}{hv_1}$, where $v_1=\frac{\hbar k_1}{m}$.
Using the orthonormality of $\chi_m(y)s$, we get,
\begin{eqnarray*}
\lefteqn{I_1(E)=} \\
&& \frac{1}{h v_1}\left[\int_{-\infty}^\infty dx\sum_m \left|f_m(x,1)\right|^2\right] \nonumber\\
& = & \frac{1}{h v_1}\left[ \int_{-\infty}^{\infty} dx \mid f_1(x,1)\mid ^2 +\int_{-\infty}^{\infty} dx \mid f_2(x,1)\mid ^2 \right.\nonumber\\
& + &\left. \int_{-\infty}^{\infty} dx \mid f_3(x,1)\mid ^2+\int_{-\infty}^{\infty} dx \mid f_4(x,1)\mid ^2+\cdots \right]\nonumber
\end{eqnarray*}
Substituting the values of $f_m(x,1)s$, we get,
\begin{eqnarray*}
  \lefteqn{I_1(E)=} \\
  && \frac{1}{hv_1} \left[ \int_{-\infty}^0 dx\left[1+\mid r_{11}\mid ^2 +2\mid r_{11}\mid \cos(2k_1 x +\phi_1)\right]\right. \nonumber \\
  & + & \int_0^\infty dx \mid t_{11}\mid^2+\int_{-\infty}^0 dx \mid r_{12}\mid ^2+\int_0^\infty dx \mid t_{12}\mid^2 \nonumber \\
 & + & \left. \frac{ \mid t_{13}\mid^2 }{\kappa_3 }+\frac{ \mid t_{14}\mid^2 }{\kappa_4 }+\cdots \right] \nonumber
 \end{eqnarray*}
Here, $r_{11}=\mid r_{11}\mid e^{-i\phi_1}$.
Note that for $m>2$, 
\begin{eqnarray*}
  \lefteqn{\int_{-\infty}^\infty dx\sum_m \left|f_m(x,1)\right|^2  } \nonumber\\
  & = & \mid t_{1m}\mid^2\left[ \int_{-\infty}^0 e^{2\kappa_mx}dx+\int_{0}^{\infty} e^{-2\kappa_mx}dx\right]  \nonumber\\
  & = & \frac{\mid t_{1m}\mid^2}{\kappa_m} \nonumber
 \end{eqnarray*}
\begin{figure}
\centering
\rotatebox{270}
{\includegraphics[width=6cm,keepaspectratio]{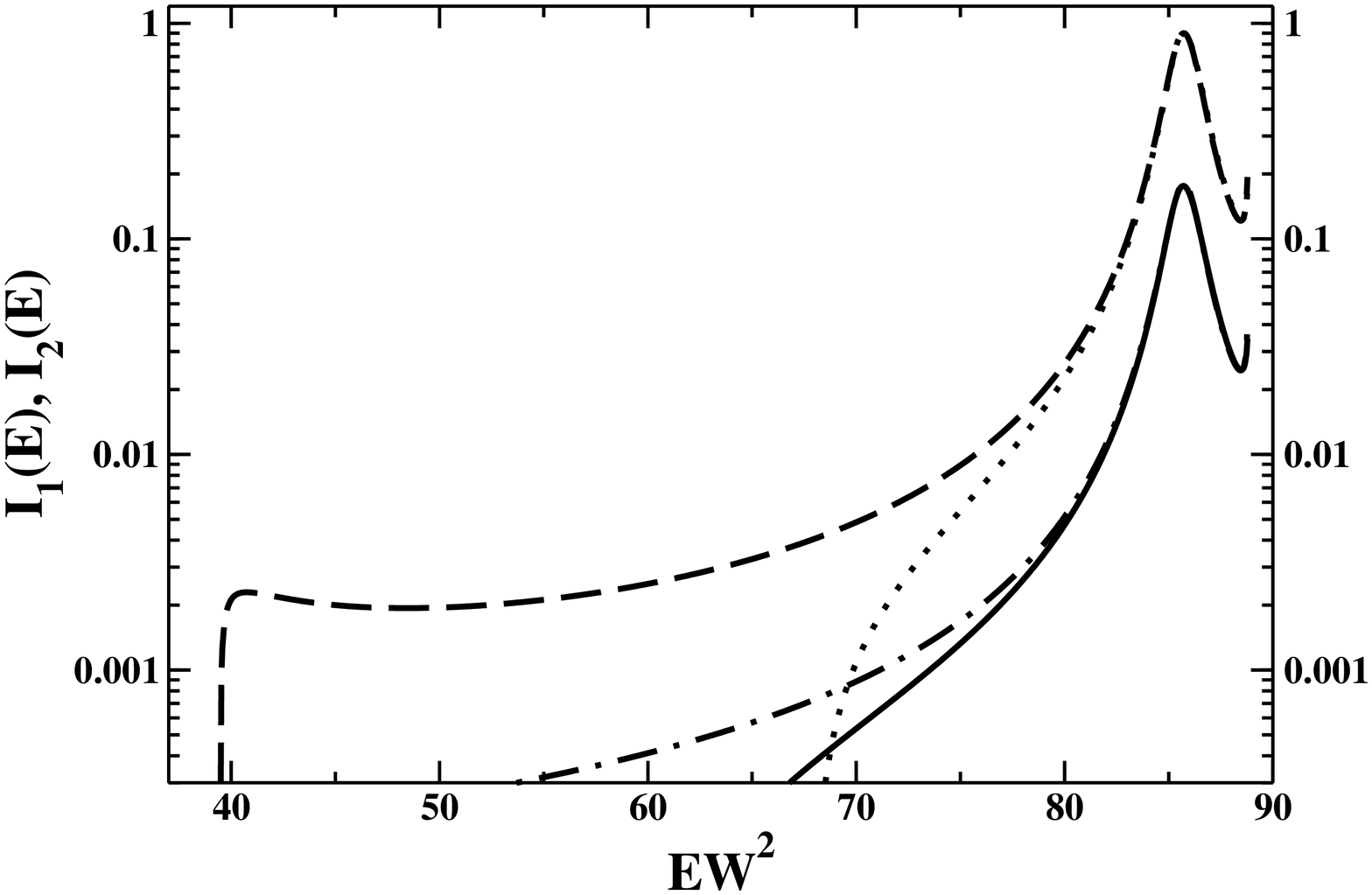}}
\caption{\label{fig4}
The plot is of injectance versus $EW^2$ for $\gamma= -13$, and $y_j=.45W$. The solid curve shows semi-classical injectance of channel 1 (i.e., incidence is along $m=1$ from left) calculated using S-matrix (Eq. (\ref{eq1})), the dot-dashed curve is the exact injectance, for the same channel calculated using internal wave function (Eq. (\ref{new4})). The dotted curve is semi-classical injectance for channel 2 from S-matrix (Eq. (\ref{new7})), the dashed curve is the exact injectance, for channel 2 (i.e., incidence is along $m=2$ from left) calculated using internal wave function (Eq. (\ref{new5})). We use $\hbar=1$, $2m=1$. The figure shows that semi-classical formula becomes exact at resonance where there is a peak in injectance.}
\end{figure}
\begin{figure}
\centering
\rotatebox{270}
{\includegraphics[width=6cm,keepaspectratio]{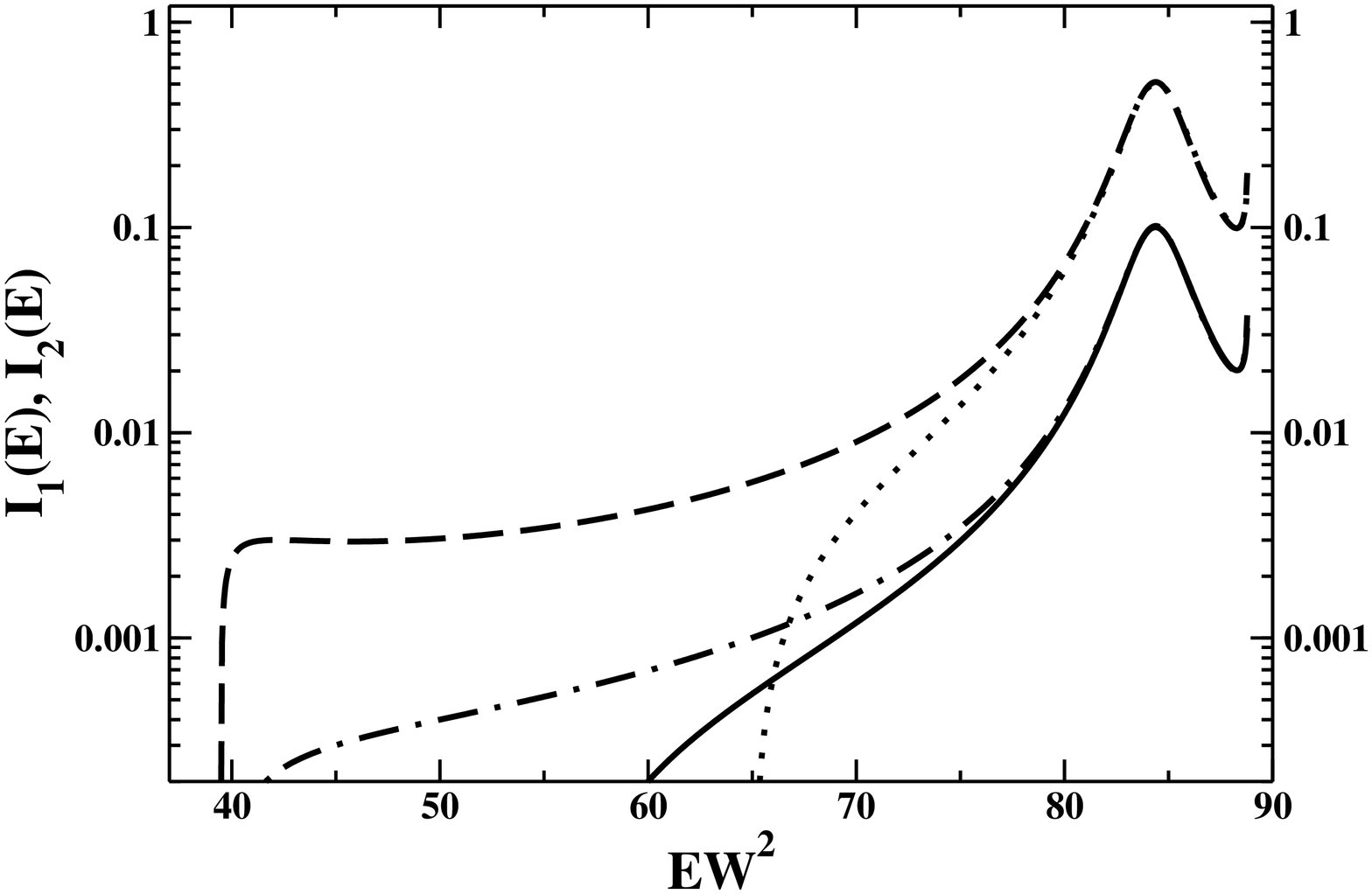}}
\caption[]{\label{fig5}
The plot is of injectance versus $EW^2$ for $\gamma= -15$, and $y_j=.45W$. The solid curve shows semi-classical injectance of channel 1 (i.e., incidence is along $m=1$ from left) calculated using S-matrix (Eq. (\ref{eq1})), the dot-dashed curve is the exact injectance, for the same channel calculated using internal wave function (Eq. (\ref{new4})). The dotted curve is semi-classical injectance for channel 2 from S-matrix (Eq. (\ref{new7})), the dashed curve is the exact injectance, for channel 2 (i.e., incidence is along $m=2$ from left) calculated using internal wave function (Eq. (\ref{new5})). We use $\hbar=1$, $2m=1$. The figure shows that semi-classical formula becomes exact at resonance where there is a peak in injectance.}
\end{figure}

Using time reversal symmetry, i.e, $r_{12}=r_{21}$ and $t_{12}=t_{21}$, we get,
\begin{eqnarray*}
I_1(E) &=& \frac{1+\mid r_{11}\mid ^2 +\mid r_{21}\mid ^2}{hv_1}\int_{-\infty} ^{0} dx \\
& + &\frac{\mid t_{11}\mid ^2 +\mid t_{21}\mid ^2}{hv_1}\int_{0} ^{\infty} dx \nonumber\\
& + &\frac{2\mid r_{11}\mid}{hv_1} \int_{-\infty} ^{0} dx \cos(2k_1x+\phi_1) \nonumber\\
& + &\frac{1}{hv_1}\left(\frac{\mid t_{13}\mid ^2 }{\kappa_3}+\frac{\mid t_{14}\mid ^2 }{\kappa_4}+\cdots \right) \nonumber
\end{eqnarray*}
Adding and subtracting the following terms,\\
\hfill\\
$\displaystyle{\frac{|t_{11}|^2}{h v_1}\int_{-\infty}^0 dx}$, 
$\displaystyle{\frac{|t_{21}|^2}{h v_1}\int_{-\infty}^0 dx}$,\\
\\ and using the fact that,
$$\frac{|t_{11}|^2}{h v_1}\int_{-\infty}^0 dx=\frac{|t_{11}|^2}{h v_1}\int_{0}^{\infty} dx$$
and, $\mid r_{11}\mid^2 +\mid r_{21}\mid^2 +\mid t_{11}\mid^2 +\mid t_{21}\mid^2=1$, we get,
\begin{eqnarray}
I_1(E) & = &\frac{1}{hv_1}\int_{-\infty}^{\infty} dx +\frac{\mid r_{11}\mid}{hv_1}\int_{-\infty}^{\infty} dx \cos(2k_1x+\phi_1)  \nonumber\\
& + &\frac{1}{hv_1}\left( \frac{\mid t_{13}\mid ^2 }{\kappa_3}+\frac{\mid t_{14}\mid ^2 }{\kappa_4}+\cdot\cdot \right)          \label{number}
\end{eqnarray}
\begin{figure}
\centering
\rotatebox{270}
{\includegraphics[width=6cm,keepaspectratio]{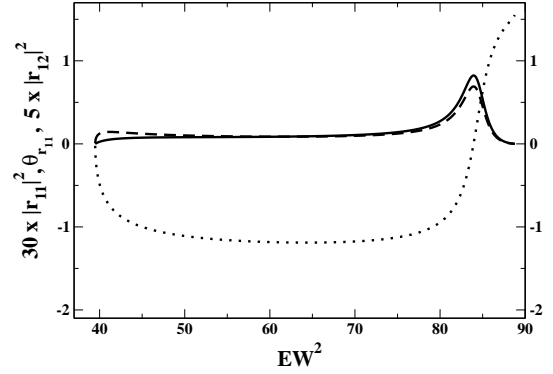}}
\caption[]{\label{fig6}
Here we are using the same parameters as that of Fig 5. The solid curve shows  $30$x$\mid r_{11}\mid^2$, the dotted curve is for $\theta_{r_{11}}$. The dashed curve is for $5$x$\mid r_{12}\mid^2$. The figure shows that at resonance, there is large fluctuation in $\mid r_{11}\mid^2$, $\theta_{r_{11}}$ and $\mid r_{12}\mid^2$. Also it shows that $\mid r_{11}\mid\neq0$ but $\sin (\theta_{r_{11}})=0$.}
\end{figure}
\begin{figure}
\centering
\rotatebox{270}
{\includegraphics[width=6cm,keepaspectratio]{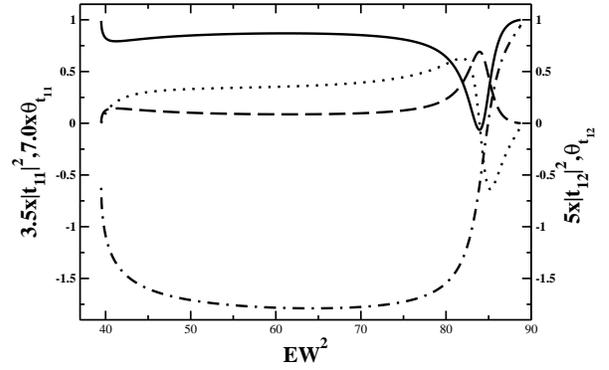}}
\caption[]{\label{fig7}
Here we are using the same parameters as that of Fig 5. The solid curve shows  $(3.5$x$\mid t_{11}\mid^2-2.5)$, the dotted curve is for $7.0$x$\theta_{t_{11}}$. The dashed curve is for $5$x$\mid t_{12}\mid^2$, and the dot-dashed curve is for $\theta_{t_{12}}-0.6$. The figure shows that at resonance, there is large fluctuation in $\mid t_{11}\mid^2$, $\theta_{t_{11}} $, $\theta_{t_{12}}$ and$\mid t_{12}\mid^2$.}
\end{figure}
$\frac{1}{hv_1}\int_{-\infty}^{\infty} dx  =I_0(E)$ is the injectance in the absence of scatterer. Again if the scattering phase shift is defined with respect to the phase shift in absence of scatterer then this term can be dropped.\\
Now,$$\int_{-\infty}^{\infty} dx \cos(2k_1x+\phi_1)=\delta(2k_1)  \cos(\phi_1)$$
As $k_1=0$ is not a propagating state contributing to transport, hence this term of Eq. (\ref{number}) reduces to zero, and we are left with,
\begin{equation}
I_1(E)=\frac{1}{hv_1}\left( \frac{\mid t_{13}\mid ^2 }{\kappa_3}+\frac{\mid t_{14}\mid ^2 }{\kappa_4}+\cdot\cdot\cdot\cdot \right)	\label{new4} 
\end{equation}
Similarly, for incidence along channel 2, one can obtain the injectance,
\begin{equation}
I_2(E)=\frac{1}{hv_2}\left( \frac{\mid t_{23}\mid ^2 }{\kappa_3}+\frac{\mid t_{24}\mid ^2 }{\kappa_4}+\cdot\cdot\cdot\cdot \right) 	\label{new5} 
\end{equation}

Note that contribution to injectance come from evanescent channels only. This is a tunneling problem because the lateral confinement makes the effective potential very large and extended in the propagating direction \cite{hua}.
In figures 4 and 5  we consider two propagating channels and show that semi-classical formula for injectance becomes exact at a resonance. 
Injectance for both the channels is shown. $I_1(E)$ is the injectance for incidence along channel 1 and $I_2(E)$ is the same for incidence along channel 2. The resonance condition is given by Eq. (\ref{bound}) and this occur at $EW^2=85.62$ in Fig.\ref{fig4} and at $EW^2=84.29$ in Fig.\ref{fig5}. At the resonant energy we can see a peak in the injectance in both channels which is a consequence of the resonance. At the peak, solid curve (semi-classical $I_1(E)$ obtained from Eq.(\ref{eq1})) coincides with dot-dashed curve (exact $I_1(E)$ obtained from Eq.(\ref{new4})), and, dotted curve (semi-classical $I_2(E)$ obtained from Eq.(\ref{new7})) coincides with dashed curve (exact $I_2(E)$ obtained from Eq.(\ref{new5})). Exactness of the semi-classical formula is counter-intuitive because at this point scattering probability and scattering phase shift show strong energy dependence, so the stationary phase approximation needed to get (\ref{eq1}), (\ref{new7}) cannot be valid. Also at this energy, $\mid r_{11}\mid\neq0$ and $\mid r_{22}\mid\neq0$, rather, $\sin(\theta_{r_{11}})=0$ and $\sin(\theta_{r_{22}})=0$ implying interference effects disappear and so correction term in (\ref{eq1}, \ref{new6}) disappear making semi-classical formula exact. All this along with the energy dependence of the scattering matrix elements can be seen in Fig.\ref{fig6} and Fig.\ref{fig7} for the parameters used in Fig.\ref{fig5}. The negative slope in the scattering phase shift $\theta_{t_{11}}$ (see dotted curve in Fig.\ref{fig6}) is due to Fano resonance \cite{yey} and another puzzling feature, explained below. 
\begin{figure}
\centering
\rotatebox{270}
{\includegraphics[width=6cm,keepaspectratio]{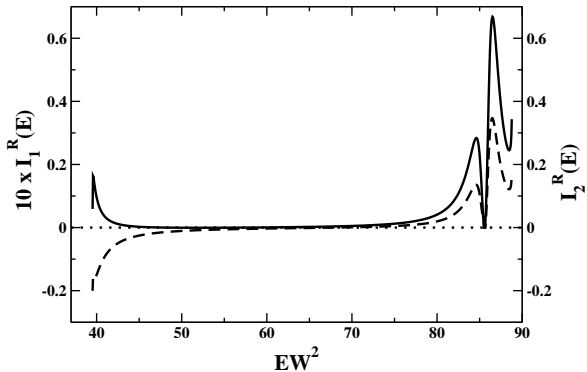}}
\caption[]{\label{fig8}
Here we are using the same parameters as in Fig.\ref{fig4}. The solid curve shows $10$x$I_1^R(E) $, the dashed line is for $I_2^R(E) $. Both of them get exactly zero at the resonance i.e., at $EW^2= 85.62$.}
\end{figure}
\begin{figure}
\centering
\rotatebox{270}
{\includegraphics[width=6cm,keepaspectratio]{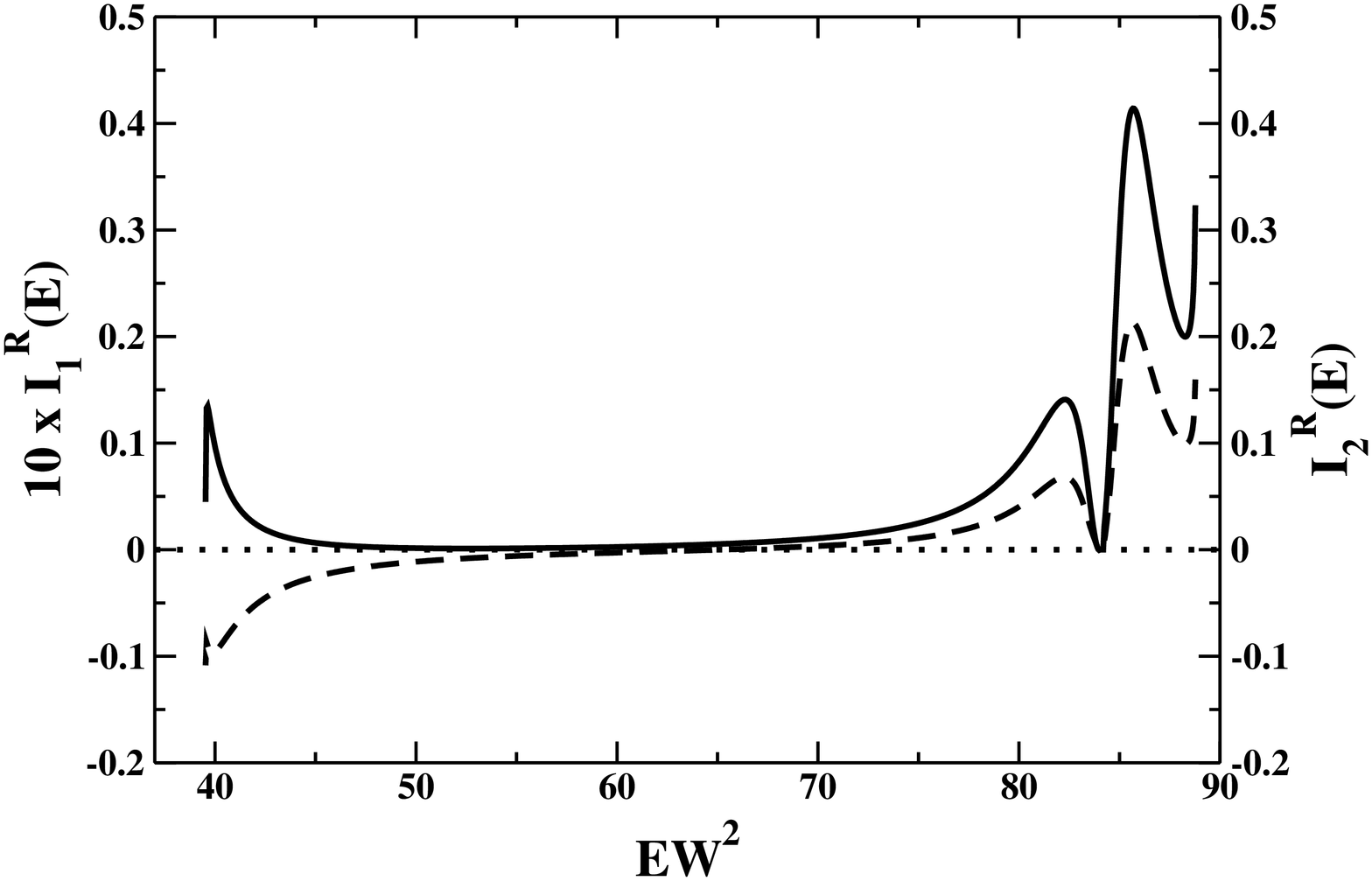}}
\caption[]{\label{fig9}
Here we are using the same parameters as in Fig.\ref{fig5}. The solid curve shows $10$x$I_1^R(E) $, the dashed line is for $I_2^R(E)$. Both of them get exactly zero at the resonance i.e., at $EW^2= 84.29$.}
\end{figure}
\begin{figure}
\centering
\rotatebox{270}
{\includegraphics[width=6cm,keepaspectratio]{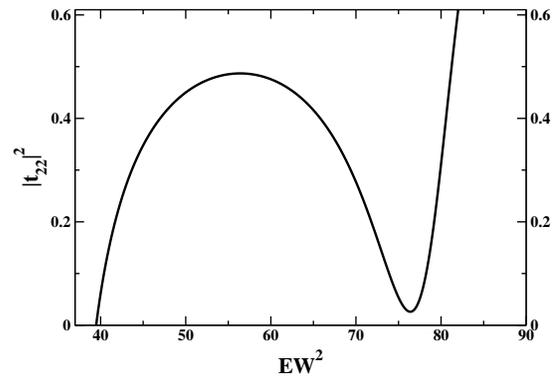}}
\caption[]{\label{fig10}
Here the energy dependence of $\mid t_{22}\mid^2$ is shown, using the same variables as in Fig.\ref{fig5}.}
\end{figure}

It can be proved that $I_{1,2}^R(E)=0$ at Fano resonance, from (\ref{new8}) and (\ref{new9}). It is shown in figures 8 and 9 for the parameters used in figures 4 and 5 respectively. As explained before that $I_{1,2}^{R}(E)$ (within a factor of $\hbar$) is a time scale associated with particles that escape to the right lead. There has been a great amount of controversy regarding, what negative or zero times mean. Note that $I_2^R(E)$ (dashed curve in Fig.\ref{fig8} and Fig.\ref{fig9}) becomes negative for $40< EW^2 <50$. So far researchers have mainly studied this low energy quantum regime [\onlinecite{smith,gsp,nus,tex,win,sala}] for observing and understanding negative time scales. In this regime the correction term is non-zero. $I_2(E)$ as defined in Eq. (\ref{new5}) (within a factor of $\hbar$) is known as dwell time and it is definitely positive. Each of the four times on the RHS of Eq. (\ref{new6}) are Wigner Smith delay time (again within a factor of $\hbar$). The controversies arise due to the lack of equality (between dashed and dotted curves in figures \ref{fig4} and \ref{fig5} ) between (\ref{new6}) and (\ref{new5}), which leaves room for different interpretations of the different times. So, negative times are also subject to interpretations. Many of these times are experimentally accessible [\onlinecite{sten}]. Only at $EW^2= 39.51$ when $k_2\rightarrow 0$, the correction term disappear because $\frac{V}{k_2}\rightarrow \infty$ (see Fig.1b)). Hence if someone can do an experiment at this energy then the negative times may be observed and experimental results are free of controversy. But at this point quite ironically, $|t_{22}|\rightarrow 0$ and so one has to wait infinitely long to see a negative time delay that is free from controversies. On the other hand at the Fano resonance ($EW^2= 85.62$ in Fig.8 and at $EW^2= 84.29$ in Fig.9), both the solid curve and the dashed curve become zero, implying there are time scales that go to zero at this energy. So the same puzzling question can be asked again, as to how can time scales be negative or zero$?$ Unlike the low energy regime, at this energy, the correction term is zero implying Wigner Smith delay times add up to give dwell time, and also transmission is finite. We show the finiteness of $\mid t_{22}\mid^2$, for same parameters as in Fig.\ref{fig5}, in Fig.\ref{fig10}. So there is no room for different interpretations and also one does not have to wait infinitely to observe a negative or zero time delay. In fact $|t_{22}|^2 \frac{d\theta_{t_{22}}}{dE} $ is strongly negative here and so if one observes the transmission to the $m=2$ channel on the left to $m=2$ channel on the right, then one can see strongly negative delays. It should not be difficult to separate such transmission processes from $t_{21}$ transmission processes, as in the former case the transmitted wave packet move much slowly (being constructed from $k_2$ instead of $k_1$). Experimentalists may try to do similar experiments in this regime. Specially as one can see, the transmissions at this energy is finite.

Study of injectance of a quantum system coupled to finite thickness leads has established many novel facts. First of all the correction term to injectance, obtained by [\onlinecite{hff}] and [\onlinecite{lev}], become zero very paradoxically in a quantum regime of a Fano resonance. For single channel leads this has been proved generally and for multichannel leads this has been shown for a delta function potential. Thus, semi-classical formulas become exact at Fano resonance. All the terms that appear in semi-classical formulas can be determined experimentally from asymptotic wave functions. Hence, although injectance strongly depend on the scattering potential, can be determined without knowing the scattering potential. Besides, the quantum interference term going to zero can unambiguously establish the possibility of negative time scales in quantum mechanics.

The authors acknowledge useful discussions with Dr. Rajesh Parwani and DST for funding this research.



\begin{thebibliography}{99}
\bibitem{hei00} Yang Ji, M. Heiblum, D. Sprinzak, D. Mahalu, Hadas Shtrikman, Science {\bf 290}, 779 (2000).
\bibitem{sch} R. Schuster  \textit{et al.}, Nature {\bf 385}, 417 (1997).
\bibitem{kob1} K. Kobayashi, H. Aikawa, S. Katsumoto and Y. Iye,
Phys. Rev. B {\bf 68}, 235304 (2003).
\bibitem{kob2} K. Kobayashi, H. Aikawa, A. Sano, S. Katsumoto and Y. Iye,
Phys. Rev. B {\bf 70}, 035319 (2004).
\bibitem{hilbertTransform} R. Englman and A. Yahalom, Phys. Rev. B {\bf 61}, 2716 (2000).
\bibitem{jpcm} P Singha Deo, J. Phys.Condens. Matter {\bf 21}, 285303 (2009).
\bibitem{lee} H. W. Lee, Phys. Rev. Lett {\bf 82}, 2358 (1999).
\bibitem{btp} M. B\"{u}ttiker, H. Thomas, and A Pretre, Z. Phys B {\bf 94}, 133 (1994);  M. B\"{u}ttiker, Pramana Journal of Physics {\bf 58}, 241 (2002).
\bibitem{merz} E. Merzbacher, Quantum Mechanics, 3rd ed. (Wiley, New York, 1997).
\bibitem{ziman} J. M. Ziman, Principles of the Theory of Solids, 2nd ed. (Cambridge University Press, Cambridge, UK, 1979).
\bibitem{dat} S. Datta, Electronic transport in mesoscopic systems, (Cambridge University Press, Cambridge, UK, 1995).
\bibitem{hff} E. H. Hauge, J. P. Falck, and T. A. Fjeldly, Phys. Rev. B {\bf 36}, 4203 (1987).
\bibitem{lev} C. R. Leavens and G. C. Aers, Phys. Rev. B {\bf 39}, 1202 (1989).
\bibitem{swa} Swarnali Bandopadhyay and P. Singha Deo,
Phys. Rev. B {\bf 68} 113301 (2003); P. Singha Deo, 
Swarnali Bandopadhyay and Sourin Das,
International Journ. of Mod. Phys. B, {\bf 16}, 2247 (2002).
\bibitem{bay} B. F. Bayman and C. J. Mehoke, Am. J. Phys. {\bf 51}, 875 (1983).
\bibitem{bag}  P. F. Bagwell, Phys. Rev. B {\bf 41}, 10354 (1990).
\bibitem{hua} H. Wu, D. W. L. Sprung and J. Martorell, Phys. Rev. B {\bf 45}, 11960 (1992).
\bibitem{yey} A. L. Yeyati and M. B\"{u}ttiker, Phys. Rev. B {\bf 62}, 7307 (2000), and references therein.
\bibitem{tan} T. Taniguchi and M. B\"{u}ttiker, Phys. Rev. B 60, {\bf 13}, 814 (1999).
\bibitem{smith} F. T. Smith, Phys. Rev. B {\bf 118}, 349 (1960).
\bibitem{gsp} V. Gasparian \textit{et al.}, Phys. Rev. A {\bf 54}, 4022 (1996).
\bibitem{nus} H. M. Nussenzvig, Phys. Rev. A {\bf 62}, 042107 (2000).
\bibitem{tex} C. Texier and M. B\"{u}ttiker, Phys. Rev. B {\bf 67}, 245410 (2003).
\bibitem{win} H. G. Winful, Phys. Rev. Lett {\bf 91}, 260401 (2003).
\bibitem{sala} Time in Quantum Mechanics, edited by J. G. Muga, R. Sala Mayato, and I. L. Egusquiza (Springer, Berlin, 2002), and references therein.
\bibitem{sten} M. D. Stenner  \textit{et al.}, Nature {\bf 425}, 695 (2003), and references therein.

\end{thebibliography}
\end{document}